\documentstyle[prl,aps,epsf]{revtex}
\newcommand{\bq}{\begin{equation}}
\newcommand{\eq}{\end{equation}}
\begin{document}
\draft
\twocolumn[\hsize\textwidth\columnwidth\hsize\csname @twocolumnfalse\endcsname
\title{
Realistic Grand Canonical Monte Carlo Surface Simulation: 
Application to Ar(111)}
\author{
Franck Celestini$^{1,2}$,
Daniele Passerone$^{1,2}$,
Furio Ercolessi$^{1,2}$, and
Erio Tosatti$^{1,2,3}$
}
\address{
$^{\rm (1)}$ International School for Advanced Studies (SISSA-ISAS),
I-34014 Trieste, Italy}
\address{
$^{\rm (2)}$ Istituto Nazionale di Fisica della Materia (INFM), Italy}
\address{
$^{\rm (3)}$ International Centre for Theoretical Physics (ICTP),
I-34014 Trieste, Italy}
\date{8 October 1997}
\maketitle

\begin{center}Preprint SISSA {\bf 133/97/CM/SS} \\
To appear on {\em Surface Science}, 1998
\end{center}
\begin{abstract}
Most realistic, off-lattice surface simulations are done 
canonically---conserving particles. 
For some applications, however,
such as studying the thermal behavior of rare gas solid 
surfaces, these constitute bad working conditions. 
Surface layer occupancies are believed to change with temperature, 
particularly at preroughening, and naturally call for a grand canonical 
approach, where particle number is controlled by a chemical potential. 
We report preliminary results of novel realistic 
grand canonical Monte Carlo simulations of the 
Lennard-Jones (LJ) fcc$(111)$ surface, believed to represent a quantitative 
model of {\it e.g.}\ Ar$(111)$.
The results are successful and highly informative for temperatures 
up to roughly $0.8\,T_m$, where clear precursor signals of 
preroughening are found. At higher temperatures, convergence to
equilibrium is hampered by large fluctuations.
\end{abstract}
\pacs{PACS numbers: 68.35.Rh, 68.35.Bs, 68.35.Md, 82.65.Dp}

\vskip2pc]

\narrowtext

\noindent
{\bf 1. Introduction}

Our quantitative understanding of thermal surface phenomena, including 
surface phase transitions, relies heavily on numerical simulations.
Classical molecular dynamics (MD) and Monte Carlo (MC) simulations 
have played and play a crucial role, describing the surface either
via lattice hamiltonians, or via continuous (off-lattice) 
classical hamiltonians, or via continuous {\em ab initio} 
hamiltonians. The continuous hamiltonians are
of course much more realistic, but they also have serious limitations.
One of these limitations, which we address here, has traditionally been
connected with strict particle conservation. It is desirable, in the 
study of many surface phenomena, to let the particle number fluctuate 
grand canonically. Lattice models permit that very naturally, but 
continuous models do not. In fact, there is so far very little experience 
of grand canonical surface studies conducted with continuous, off-lattice 
MD and MC simulation.
In this paper we report on groundwork, plus some initial successes,
in implementing a grand canonical Monte Carlo (GCMC) surface simulation 
based on a continuous classical hamiltonian, in particular the
Lennard-Jones fcc$(111)$ surface, chosen as a model for rare gas
surface.
 
Rare gas solid surfaces and films have provided an important testing 
ground for several surface phase transitions for over two decades. 
Surface melting, roughening, and recently preroughening have been
studied and characterized at the free solid-vapor interface. Layering
transitions of thin rare gas film on smooth substrates have given rise 
to a vast literature. The discovery of reentrant layering of rare gases
on substrates has led to a debate \cite{debat} over a possible explanation
in terms of preroughening, as suggested by RSOS models \cite{weich}, 
versus a melting-solidification interplay  \cite{philips} as  argued from 
canonical simulations with a continuous potential. In real life, 
the true rare gas solid 
surface begins to develop diffusion phenomena,
which are totally absent in lattice RSOS models, above roughly
$(1/2)\,T_m$. However, 
the lattice models allow a much more thorough statistical mechanical
understanding and classification of possible surface phases. In particular,
they imply the possible existence of a preroughening (PR) transition, leading 
to a so called Disordered Ordered Flat (DOF) phase. 
At $T_{\rm PR}$, among other 
things, the surface occupancy exhibits a transition from close to a full
monolayer $(T<T_{\rm PR})$ to close to a half monolayer $(T>T_{\rm PR})$.
This kind of transition cannot be easily simulated  with a fixed particle
number. In fact it has been shown elsewhere \cite{prestip,jayanthi} 
that a fixed particle number will lead, at and above $T_{\rm PR}$, to a kind 
of phase separation into two neighboring DOF phases. In other words, 
that surface will not even remain macroscopically homogeneous. Classical
off-lattice GCMC \cite{Rowley}  has so far be succesfully applied to simulate
different systems such as capillary condensation in nanopores \cite{Peterson}. 
However difficulties appear when one tries to simulate a denser system 
\cite{Mezei} and the classical GCMC technique requires modifications. In the
following we shall describe our own implementation.
\vspace{5mm}

\noindent
{\bf 2. Grand canonical Monte Carlo surface simulation: technical details.}

Our implementation of GCMC is based on well known techniques \cite{Rowley},
with modifications to improve the sampling efficiency in a system
with a free surface.  In particular, taking inspiration from an idea 
previously used for bulk liquids \cite{Mezei},
we have restricted attempts to create or destroy particles to a region 
near the surface, where they are much more likely to be accepted.

We have considered three different types of Monte Carlo moves:

(a) small thermal displacements of individual atoms, with magnitudes 
$\delta r$ adjusted with $T$ in order to obtain an optimal acceptance
rate of $50 \%$, and random direction;

(b) large lateral displacements of surface atoms;

(c) destruction of an existing particle, or creation of a new particle.
Since deep in the bulk the acceptance rates will generally be 
exceedingly small, we restrict attempts to within a region of thickness  
$d = na$, where $a$ is the spacing between two consecutive bulk lattice 
layers, and $n$ is typically 6. This region is centered on the outer layer 
of the system simulated.

Moves of type (b) were thought at first to be important for establishing
surface diffusivity. It was later seen, however, that this is not really so
for the LJ surface. Therefore, they were eventually omitted. 
Clearly, only moves of type (c) are grand canonical. Thermal moves (a)
must be by far the most frequent, for equilibration. We find that a 
large number of such moves is necessary, in the temperature range 
investigated, between a grand canonical move (c) and the next. 
We have taken the probability of attempting a move (c) to be
$\alpha_c = \alpha_d = \alpha$ (respectively for creation and destruction), 
and that of attempting a move (a) to be $\alpha_m = 1 - \alpha$,  
with $\alpha$  very small, its actual value depending on the ratio between the
number of atoms in the outer layers and
the total number of atoms, but roughly of the order of $0.5 \times 10^{-3}$.

The acceptance probability of creation and destruction, $p_c$ and $p_d$
have then been finally normalized in the form \cite{Rowley}: 
\begin{equation}
p_c = \min\left[1,\frac{V}{(N+1)\Lambda^3}
\exp\left( \frac{\mu + \delta U}{k_B T} \right) \right]
\end{equation}
\begin{equation}
p_d = \min\left[1,\frac{N \Lambda^3}{V}
\exp\left( \frac{-\mu + \delta U}{k_B T} \right) \right]
\end{equation}
where $N$ is the number of atoms present in the creation/destruction region,
$V$ its volume,
$\Lambda = [h^2/(2\pi m k_B T)]^{1/2}$ is the thermal de Broglie wavelength,
and $\delta U = U_a - U_b$ is the difference between the
total energy $U_a$ after the trial move (creation or destruction) and
the total energy $U_b$ before it.
In order to satisfy detailed balance $N$ must be 
reajusted at each step of the simulation. 

We  simulated in this way the free fcc$(111)$ solid-vapor interface of  
a rare gas, particularly Ar. The interatomic forces were described by a 
(12,6) Lennard-Jones 
potential truncated at $2.5\,\sigma$. Our system consisted of 
a 15-layer slab, with
periodic boundary conditions along the $x$ and $y$ direction in the interface
plane. Three bottom atomic layers were kept 
frozen, while 12 layers of 480 atoms each were free to evolve. 
The lateral box size was rigid, but readjusted at each temperature according 
to the thermal expansion coefficient of this potential, obtained from separate
bulk simulations \cite{furio_priv}. We considered a grid of temperatures above
$0.46 \epsilon$, as there is very little action below (the bulk melting 
temperature is $T_B \simeq 0.7 \epsilon$). For each  temperature
we first found the equilibrium value of the effective chemical potential 
$\mu_\circ(T)$. This was done by trial and error, starting from an 
arbitrary value, and 
changing it until the particle number remained as stationary as possible.
With $\mu_\circ(T)$ so determined we made long equilibration runs looking for  
stabilization of both the total energy and the number of particles in the 
system. Generally half a million of Monte-Carlo (MC) steps (in the usual
sense, i.e., one step attempting to move each particle on average once)
were sufficient to reach reasonable equilibrium . Here we encountered our 
main problem, which 
is that this procedure is increasingly less convergent, and the 
resulting surface less stable, as $T$ increases.
In practice, we found it impossible 
to stabilize the surface against exceeedingly large fluctuations 
above $T=0.56 \epsilon$ ($\simeq 0.80\,T_m$). This 
constitutes an important drawback, because the interesting phase transitions
of the surface are believed to lie just above this temperature, {\it i.e.,} 
preroughening at  $0.83\,T_m$ and roughening at $0.95\,T_m$ \cite{jayanthi}.
On the other hand, a further reduction in the relative creation/destruction
rate $\alpha$, necessary to obtain grand-canonical equilibrium at these
higher temperatures, would quickly
become too costly, as would the alternative possibility of creating/destructing
atoms directly in the vapor phase.
Below $0.80\,T_m$, where our equilibration worked,  typically $30$ to $50$ 
uncorrelated configurations were subsequently generated from another half
a million MC steps run, and finally analysed.
\vspace{5mm}

\noindent
{\bf 3. Results and discussion}

\begin{figure}
\vspace*{0.5cm}
\epsfxsize=10cm
\epsfbox{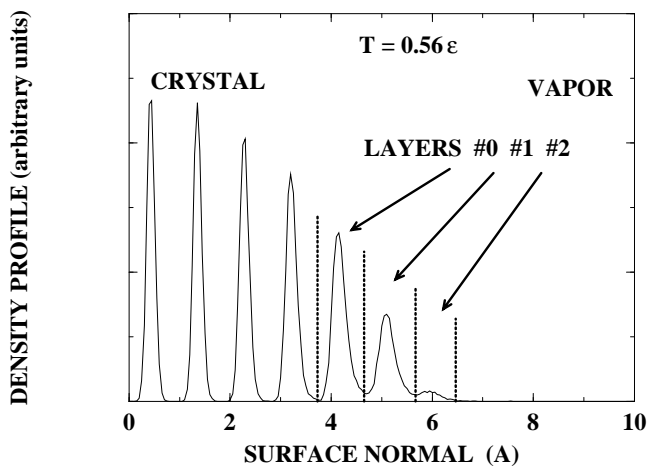}
\vspace*{-6.5cm}
\caption{
($xy$)-averaged  density profile of the simulated Ar fcc(111) 
surface in grand canonical equilibrium at $T = 0.56 \epsilon$, or 
$T = 0.8\,T_m$, plotted along the surface normal. The sub-monolayer
coverage in layer 1 is the main new result, in comparison with
previous canonical simulations.
}
\label{fig:profile}
\end{figure}

\begin{figure}
\vspace*{0cm}
\epsfxsize=10cm
\epsfbox{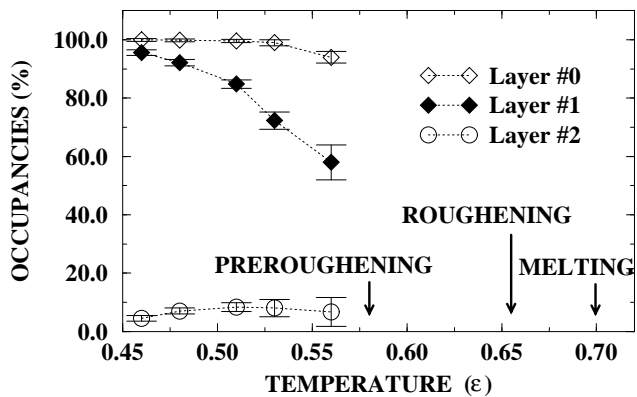}
\vspace*{-7cm}
\caption{
Occupancies of the three outermost layers (fraction of particles
contained in the bins indicated in fig.\ \ref{fig:profile}) 
as a function of temperature. 
Note the gradually decreasing occupancy of layer 1, extrapolating to the
preroughening value of 1/2 close to $T = 0.58 \epsilon$.
}
\label{fig:occupancies}
\end{figure}

The typical ($xy$)-averaged density profile of the simulated 
grand canonical Ar(111) surface (really solid-vapor interface, but the 
vapor density is tiny) is shown in fig.\ \ref{fig:profile}.
Layers are clearly visible, and the outermost
ones are labeled 2 (``adatom layer''),  1 (``surface layer''), and 0
(``subsurface layer''). We conventionally associate an atom with a given layer
$n$, if it falls within the bin $n$, as indicated. 
Taking the ratio of the number of atoms in layer $n$, $N_{n}$ to the number
$N_B$ of atoms in a bulk layer, we obtain a conventional layer occupancy
$o_{n}= N_{n}/N_B$.  Occupancies of the three outermost layers  are shown
in fig.\ \ref{fig:occupancies} as a function of $T$. 
At low temperatures,  the layers $0$ and $1$ were almost full, with only 
a few percent of vacancies, with corresponding few adatoms in layer $2$. 
As $T$ increased, layer $1$ lost atoms very rapidly, 
reaching an occupancy $o_{1}\simeq 55\%$ at $T = 0.56 \epsilon$.

It is important at this point to compare the properties of this 
grand canonical Lennard-Jones (111) surface with those of exactly 
the same surface, simulated under identical physical conditions except for
the canonical, particle-conserving constraint. The results obtained
in our earlier, very extensive MD simulations 
\cite{jayanthi}  were quite different. 
By starting with an integer number of layers, and at 
the very same $T = 0.56 \epsilon$, for example, we had obtained canonically
$o_{2}\simeq 100\%$, $o_{1}\simeq 90\%$, $o_{0}\simeq 10\%$, against
the present grand canonical values $o_{2}\simeq 94\%$, 
$o_{1}\simeq 55\%$, $o_{0}\simeq 8\%$.  Grand canonically, the ``total surface 
population'' $(o_{2}+ o_{1}+ o_{0}-1)$  appears to change continuously 
with temperature, going from nearly $100\%$ at $0.65\,T_m$  down to $57\%$ 
at $0.8\,T_m$.  This kind of gradual change was not predicted by the 
less realistic solid-on-solid models \cite{prestip,prestip2}. 
  
Extrapolating, these data strongly suggests that at 
$T\simeq 0.58 \epsilon$  the surface spontaneously tends to half 
occupancy, with about  $8\%$ adatoms  and $6\%$ subsurface vacancies. 
This temperature is in excellent agreement with the experimentally observed
temperature for the onset of reentrant layering, 
$T = 69\,{\rm K} \simeq 0.83\,T_m$ \cite{hess}. 
The coincidence with the preroughening temperature
obtained from canonical simulations is perfect. The half occupancy also
agrees with theoretical predictions and with solid-on-solid simulations of the
disordered flat (DOF) state, realized at and above 
preroughening \cite{prestip2,dennijs}. Therefore, the present
grand canonical realistic simulation data strongly confirm the 
occurrence of preroughening and of a DOF phase 
on the free surface of Ar(111) at $T_{\rm PR} = 0.83\,T_m$.
     
\begin{figure}
\vspace*{-1cm}
\epsfxsize=9cm
\epsfbox{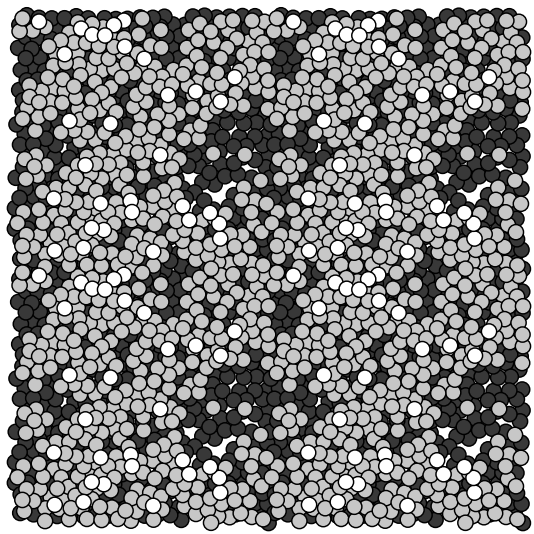}
\vspace*{-3cm}
\caption{
Top view (snapshot) of the three outer layers of simulated
grand canonical Ar(111) at $0.8\,T_m$. This surface is nearly
``disordered flat'' (DOF). Atoms in layers $0$, $1$ and $2$ ,
identified through their $z$-coordinate as in fig.\ \ref{fig:profile}, are 
respectively black, grey and white. For convenience, four adjacent 
cells are shown. The nearly half occupancy
of the surface layer (layer $1$) is realized through large islands 
and large craters.
}
\label{fig:topview}
\end{figure}
Fig.\ \ref{fig:topview}  presents a top snapshot of the grand canonical
surface at  $T = 0.56 \epsilon$, the closest we can get to $T_{\rm PR}$.
This is, we believe, the first available illustration of what a rare
gas surface really looks like at preroughening. The main features
to be noted are the large islands and craters in the first (gray) layer.
We did not try yet to examine the height-height correlations
of this surface, which would have been very interesting, 
particularly to check whether they are large as expected
by our predicted divergence at  $T_{\rm PR}$\cite{prestip}. In order to 
do that properly, we would need finite-size scaling, which is computationally
too demanding.
\vspace{5mm}

\noindent
{\bf 4. Conclusions}

In conclusion, we have reported preliminary results on the microscopic
nature of a rare gas solid-vapor interface, as obtained by means of
a novel grand canonical Monte Carlo simulation. In spite of difficulties
encountered with stabilizing the surface at higher temperatures, we
have succeeded in describing what appears to be a well equilibrated
surface in an interesting temperature range. We have obtained a clear 
indication that preroughening is indeed taking
place under conditions very close to those where reentrant layering
was observed on Ar(111), and where DOF phase separation was found by
canonical MD \cite{jayanthi}. We expect that, with the improvements 
which experience should soon bring about, grand canonical
surface simulations should become {\it the} standard approach in the 
near future. 
\vspace{5mm}

\begin{sloppypar}
It is a pleasure to thank S. Prestipino, G. Santoro, and C.~S. Jayanthi 
for constructive discussions. We acknowledge partial support from the 
European Commission under contracts ERBCHRXCT920062 and ERBCHRXCT930342,
and INFM under PRA LOTUS. Work at SISSA by F.\ C.\ is under European 
Commission sponsorship, contract ERBCHBGCT940636.
\end{sloppypar}


\end{document}